\documentclass[prd,showpacs,amsmath,amssymb,superscriptaddress,nofootinbib,showkeys]{revtex4}
\usepackage{amssymb}
\usepackage{amsmath}
\usepackage[cp1251]{inputenc}
\usepackage[T2A]{fontenc}
\usepackage{graphicx}
\usepackage{floatflt}

\DeclareGraphicsExtensions{.pdf,.eps,.png,.jpg}

\begin{document}
\title{Models of exponential and power-law acceleration of the Universe in Horndeski theory without ghosts and Laplace instabilities }

\author{Ruslan K. Muharlyamov}
\email{rmukhar@mail.ru} \affiliation{Department of General
Relativity and Gravitation, Institute of Physics, Kazan Federal
University, Kremlevskaya str. 18, Kazan 420008, Russia}

\author{Tatiana N. Pankratyeva}
\email{ghjkl.15@list.ru} \affiliation{Department of Higher
Mathematics, Kazan State Power Engineering University,
Krasnoselskaya str. 51, Kazan 420066, Russia}

\author{Shehabaldeen O.A. Bashir}
\email{shehapbashir@gmail.com} \affiliation{Department of General
Relativity and Gravitation, Institute of Physics, Kazan Federal
University, Kremlevskaya str. 18, Kazan 420008, Russia; Department of physics, Faculty of Science, University of Khartoum, Khartoum, Sudan}

\begin{abstract}
We present a  designer method for flat Friedman–Robertson Walker space-times within the framework of Horndeski's scalar-tensor theory.
In this method, as ansatzes, we use the following assumptions. Firstly, we set the propagation speeds $c_T$, $c_S$ in the tensor and scalar sectors of perturbations, the tensor-to-scalar ratio $r$
and other perturbation parameters to be constant. Following the observational data, we believe that $1-3\cdot 10^{-15}<c_T<1$, $0<r<0.1$ (APJ Lett.
-- 2017. -- V.848.-- P.L13, arXiv:1807.06209). This ansatz allows us to exclude ghosts and Laplace instabilities at the initial stage.
Secondly, we choose the dependence of the Hubble parameter on the derivative of the scalar field $H(\dot{\phi})$.  Since the characteristics of perturbations $c_T$, $c_S$ and others are initially expressed through $H$, $\dot{\phi}$ and the Horndeski potentials, the ansatzes with the gravity equations give a system of differential equations for the scalar field and the unknown Horndeski potentials. Knowledge $H(\dot{\phi})$ and $\dot{\phi}(t)$ gives the law of the Universe expansion $a(t)$. As a result, one can find subclasses of  Horndeski's scalar-tensor theory within which cosmological solutions exist without ghosts and Laplace instabilities. As an example, we found subclasses of Horndeski's theory for exponential and power-law inflation models of the Universe without  pathologies.

\end{abstract}

\pacs{04.50.Kd}

\keywords{Horndeski theory; dark
energy theory; cosmological perturbation theory}

\maketitle

\section{Introduction}

The modern database of observational cosmology leads to questions that don’t have satisfactory and full explanations within the framework of the general theory of relativity (GR) and the Standard Model of elementary particles. A possible way out of this situation is a modification of GR. The modification leads to additional degrees of freedom in field theories, which gives new mechanisms for modeling physical processes and the ability to regulate them according to observational data. From this position, the Horndeski gravity (HG) \cite{Horndeski} has «flexibility», since it has a wide phenomenology -- its Lagrangian contains four arbitrary functions. HG gives differential equations containing derivatives of no higher than the second order. The structure of the Lagrangian of this theory and, as a consequence, the restriction on the order of derivatives in the field equations allow us to avoid the Ostrogradsky instability. We use the following parametrization of the action density for HG \cite{Kobayashi1}:
\begin{eqnarray}\label{lagr1} L_H=\sqrt{-g}\Big(\mathcal{L}_2+\mathcal{L}_3+\mathcal{L}_4+\mathcal{L}_5\Big) \,,\end{eqnarray}
$$\mathcal{L}_2 = G_2(\phi,X)\,,\, \mathcal{L}_3 = -
G_3(\phi,X)\Box\phi\,,$$
$$\mathcal{L}_4 = G_{4}(\phi,X) R +G_{4X}(\phi,X) \left[ (\square
\phi )^{2}-(\nabla_\mu \nabla_\nu \phi)^2  \right]\,,$$
\begin{equation} \mathcal{L}_5 = G_{5}(\phi,X) G_{\mu\nu}\,\nabla^\mu \nabla^\nu
\phi -\frac{1}{6}G_{5X}   \left[\left( \Box \phi \right)^3 -3 \Box
\phi (\nabla_\mu \nabla_\nu \phi)^2 + 2\left(\nabla_\mu \nabla_\nu
\phi \right)^3 \right], \label{lagr2}
\end{equation}
where $g$ is the determinant of metric tensor
$g_{\mu\nu}$; $R$ is the Ricci scalar and $G_{\mu\nu}$ is the
Einstein tensor; the factors $G_{i}$ ($i=2,3,4,5$) are arbitrary
functions of the scalar field $\phi$ and the canonical kinetic
term, $X=-\frac{1}{2}\nabla^\mu\phi \nabla_\mu\phi$. We consider
the definitions $G_{iX}\equiv \partial G_i/\partial X$,
 $(\nabla_\mu \nabla_\nu \phi)^2\equiv\nabla_\mu \nabla_\nu \phi
\,\nabla^\nu \nabla^\mu \phi$, and $\left(\nabla_\mu \nabla_\nu
\phi \right)^3\equiv \nabla_\mu \nabla_\nu \phi  \,\nabla^\nu
\nabla^\rho \phi \, \nabla_\rho \phi  \nabla^\mu \phi$.
 Here we will limit ourselves to theory
\begin{equation}\label{g4234} G_4=\mu/2\,\, (\mu=M^2_{PL}\,\, \text{is the Planck mass squared}), \,\,G_2(X,\phi),\,\, G_3(X,\phi),\,\, G_5(X).\end{equation}
Subclass of HG (\ref{g4234}) contains special cases: $G_4=\mu/2$, $G_2=G_3=G_5=0$ -- this corresponds to the vacuum general relativity; $G_4=\mu/2$, $G_2=X-V(\phi)$, $G_3=G_5=0$ -- the general relativity with the conventional scalar field; $G_4=\mu/2$, $G_2(X,\phi)$, $G_3=G_5=0$ -- the K-essence theory; $G_4=\mu/2$, $G_2(X,\phi)$, $G_3(X,\phi)$, $G_5=0$ -- the KGB theory. The function $G_5(X,\phi)$ gives a non-minimal kinetic coupling to the spacetime curvature \cite{Gao, Granda}, which may appear in some Kaluza-Klein theories \cite{Shafi1,Shafi2}. In paper \cite{Amendola}  it is shown that the scalar-tensor theory with non-minimal kinetic coupling is not conformally equivalent to Einstein's theory with minimally coupled scalar field with the potential, and therefore its consequences may have qualitatively new features \cite{Sush2012,Sush2023}.
 The potential $G_5(X,\phi)$ leads to a deviation of the speed of gravitational waves $c_T$  from the speed of light. The gravitational-wave event GW170817 showed that this deviation is small for the redshift $z<0.009$. \cite{LIGO}. This places tight constraints on dark energy models constructed in the framework of modified gravitational theories. For example, the paper \cite{Kase} presents a review of Horndeski models surviving constraints from GW170817.  HG is able to fit other observational data \cite{Benjamin}, \cite{Gu}.  In work \cite{Benjamin} authors presented constraints on HG from a cosmic shear analysis of the final data release of the Kilo-Degree Survey in combination with DESI measurements of baryon acoustic oscillations, eBOSS observations of redshift space distortions, and cosmic microwave background  anisotropies from Planck.   In papers \cite{Eric18,Eric20}, the subclasses of HG, which are called No Slip Gravity, No Run Gravity, are studied for their correspondence to observational data.

Modern high-quality instruments make it possible to obtain a wealth of observational cosmological data \cite{DESI1}. The observations of the Wilkinson Microwave Anisotropy Probe (WMAP) \cite{Hinshaw}, Planck satellites \cite{PAR} and the Dark Energy Spectroscopic Instrument (DESI) \cite{DESI2} say that the modern Universe is isotropic to a certain extent.
 Taking this fact into account, we will take as a basis an isotropic space-time, namely flat Friedman–Robertson–Walker (FRW) space-times. As is known, the standard Big Bang theory model must contain the necessary stage of inflationary expansion of space-time. We are focused on searching solutions with inflation.

The standard solution algorithm involves choosing potentials $G_i(X,\phi)$, obtaining a dependence for the scale factor $a(t)$, the scalar field $\phi(t)$, and analyzing them for compliance with the observed data. The search result for a solution is not predictable. Another way is the {\bf designer} method. The work by Ellis and Madsen (1991) \cite{Ellise} was one of the first to consider "the inverse problem"\,within the framework of cosmology. The scale factor served as an ansatz. Next, from the cosmological equations, the potential $V(\phi)$ and the scalar field $\phi$ were found as  functions of time.  This method has been developed in many works, for example \cite{Chervon1}, \cite{Chervon2}. There are various variations of the initial assumption (ansatz) in scientific papers. Of greatest interest is the designer based on observational data. Starobinsky \cite{Starobinskyaadf} used the growth factor of inhomogeneities in model with a scalar field, baryons and dark matter; Huterer and Turner (1999)\cite{Huterer}, Nakamura and Chiba (1999) \cite{Nakamura}  used the luminosity data. In works \cite{Muharlyamov2}, \cite{Muharlyamov4} the authors started from the anisotropic properties of space-time.  The  designer method has found wide application in HG \cite{Chervon3,Gao1,Kennedy,Arjona,Bernardo1,Bernardo2,Muharlyamov1, Muharlyamov2,Muharlyamov4,Muharlyamov3, Muharlyamov5}.  The rich phenomenology of HG allows solving several problems in one model.   One such problem is the question of stability.
The experience of research by various authors shows that this is a non-trivial problem \cite{Muharlyamov1}, \cite{DeFelice1}, \cite{SushkovStar}.  Authors \cite{Charmousis, Bernardo3, SAppleby} solve the cosmological constant problem and other problems using the screening method. A similar problem is solved within the framework of teleparallel
Horndeski theory \cite{Bernardo4,Bernardo5}.

In this article, for case (\ref{g4234}) we develop a  designer algorithm different from the previous ones \cite{Muharlyamov1, Muharlyamov2, Muharlyamov4, Muharlyamov3, Muharlyamov5}.
  One of the purposes of the algorithm is to eliminate ghosts and Laplace instabilities at the initial stage. To do this we set the propagation speeds $c_T$, $c_S$ in the tensor and scalar sectors of perturbations, the tensor-to-scalar ratio $r$ and other perturbation parameters to be constant. By choosing constants, we eliminate physical instabilities on all time scales and can satisfy observational data: $1-3\cdot 10^{-15}<c_T<1$, $0<r<0.1$ \cite{LIGO}, \cite{Aghanim}. Initially, the characteristics of the disturbances $c_T$, $c_S$ and others are initially expressed through $H$, $\dot{\phi}$ and the Horndeski potentials. If we make an assumption about dependence $H(\dot{\phi})$, then the gravitational equations and the conditions of constant characteristics of perturbations will give a system of differential equations for  the scalar field and the unknown Horndeski potentials. Knowledge $H(\dot{\phi})$ and $\dot{\phi}(t)$ gives the law of the Universe expansion $a(t)$.  This algorithm allows generating subclasses of HG without physical instabilities in cosmological solutions.  As an example, we found subclasses of HG ($G_2(X,\phi)$, $G_3(X,\phi)$, $G_5(X)$) for exponential and power-law inflation models of the Universe without  pathologies.

\section{Field equations and the stability conditions}\label{sec2}

 In this section we will write out the gravitational equations and the stability conditions for the cosmological background for the action density (\ref{g4234}).

We take as a basis the metric
\begin{equation}
 ds^2 = -dt^2 + a^2(t)(dx^2 +dy^2 + dz^2), \label{met0}
\end{equation}
which corresponds to Friedman's flat world.
Denoting $H =\dot{a}/a$ the Hubble parameter, the non-trivial
gravitational equations for (\ref{lagr1}), (\ref{g4234}) are
$$3\mu H^2
=-G_2
+\dot{\phi}^2G_{2X}+3G_{3X}H\dot{\phi}^3
-G_{3\phi}\dot{\phi}^2+$$\begin{eqnarray}\label{gr00}+\dot{\phi}^3(5G_{5X}
+G_{5XX}\dot{\phi}^2)H^3,\end{eqnarray}
$$\mu\big(2\dot{H}+3H^2\big)=-G_2+G_{3\phi}\dot{\phi}^2
+G_{3X}\dot{\phi}^2\ddot{\phi}+$$\begin{eqnarray}\label{gr11}+\frac{d}{dt}\left[G_{5X}\dot{\phi}^3
H^2\right]
+2G_{5X}\dot{\phi}^3 H^3\,,\end{eqnarray}
where $G_{3\phi}\equiv \partial G_3/\partial \phi$.

 One of the important criteria for the viability of the cosmological background is stability. We will exclude different physical instabilities for scalar and tensor perturbations:
1) Laplacian instabilities, which occur when the square of the propagation speed of background perturbations
becomes negative; 2)  ghost instabilities due to a wrong sign of
the kinetic term of perturbations. In the framework of HG, the conditions for the avoidance of Laplace instabilities for scalar and tensor perturbations have the form \cite{DeFelice1}, respectively:
\begin{equation}c_S^2\equiv\frac{3(2w_1^2w_2H-w^2_2w_4+4w_1w_2\dot w_1-2w_1^2\dot w_2)}{w_1(4w_1w_3+9w_2^2)}\geq 0\,,\label{cS}
\end{equation}
\begin{equation}c^2_{T}\equiv\frac{w_4}{w_1}\geq 0\,.\label{csT}
\end{equation}
 To exclude ghosts, the following inequalities must be satisfied:
\begin{equation}
Q_{S}\equiv\frac{w_{1}(4w_{1}w_{3}+9w_{2}^{2})}{3w_{2}^{2}}>0\,,
\label{QS}
\end{equation}
\begin{equation}Q_T\equiv\frac{w_1}{4}>0,\label{QT}
\end{equation}
where
\begin{equation}w_{1}  \equiv  2\,(G_{{4}}-2\,
XG_{{4X}})-2X\,(G_{{5X}}{\dot{\phi}}H-G_{{5\phi}}) \,,
\label{w1def}\end{equation}
$$w_{2}  \equiv  -2\, G_{{3X}}X\dot{\phi}+4\,
G_{{4}}H-16\,{X}^{2}G_{{4{ XX}}}H+4(\dot{\phi}G_{ {4\phi X}}-4H\,
G_{{4X}})X+$$
\begin{equation} +2\, G_{{4\phi}}\dot{\phi}+8\,{X}^{2}HG_{{5\phi X}}+2H\, X\,(6G_{{5\phi}}-5\,
G_{{5X}}\dot{\phi}{H})-4G_{{5{
XX}}}{\dot{\phi}}X^{2}{H}^{2},\end{equation}
$$w_{3}  \equiv  3\, X(G_{2{X}}+2\, XG_{{
2XX}})+6X(3X\dot{\phi}HG_{{3{ XX}}}-G_{{3\phi X}}
X-G_{{3\phi}}+6\, H\dot{\phi}G_{{3X}})+$$
$$+18\, H(4\, H{X}^{3}G_{{4{  XXX}}}-HG_{{4}}-5\, X\dot{\phi}G_{{4\phi
X}}-G_{{4\phi}}\dot{ \phi}+7\, HG_{{4X}}X+$$
$$+16\, H{X}^{2}G_{{4{
XX}}}-2\,{X}^{2}\dot{\phi}G_{{4\phi{
XX}}})+6{H}^{2}X(2\, H\dot{\phi}G_{{5{
XXX}}}{X}^{2}-$$
\begin{equation}-6\,{X}^{2}G_{{5\phi{  XX}}}+13XH\dot{\phi}G_{{5{
XX}}}-27G_{{5\phi
X}}X+15\, H\dot{\phi}G_{{5X}}-18G_{{5\phi}}),\end{equation}
\begin{equation}
w_{4}  \equiv  2G_{4}-2XG_{5\phi}-2XG_{5X}\ddot{\phi}~.
\end{equation}

Here we will assume
\begin{equation}\label{g3} G_3(X,\phi)=G^{(x)}(X)+G(\phi).\end{equation}
Substituting (\ref{g3}) into the system (\ref{gr00})-(\ref{gr11}), we get
$$3\mu H^2
=-G_2
+\dot{\phi}^2G_{2X}+3G^{(x)}_{X}H\dot{\phi}^3
-G_{\phi}\dot{\phi}^2+$$\begin{eqnarray}\label{1gr00}+\dot{\phi}^3(5G_{5X}
+G_{5XX}\dot{\phi}^2)H^3,\end{eqnarray}
$$\mu\big(2\dot{H}+3H^2\big)=-G_2+G_{\phi}\dot{\phi}^2
+G^{(x)}_{X}\dot{\phi}^2\ddot{\phi}+$$\begin{eqnarray}\label{2gr11}+\frac{d}{dt}\left[G_{5X}\dot{\phi}^3
H^2\right]
+2G_{5X}\dot{\phi}^3 H^3\,.\end{eqnarray}
Taking into account (\ref{g4234}) and (\ref{g3}), functions $w_i$ will take the form:
\begin{equation}\label{w1w4} w_1=\mu-2G_{5X}X\dot{\phi}H,\,\, w_4=\mu-2G_{5X}X\ddot{\phi},\end{equation}
\begin{equation}\label{w2}w_2=2\mu H-2G^{(x)}_{X}X\dot{\phi}-2H^2\dot{\phi}X(5G_{5X}+2G_{5XX}X),\end{equation}
$$w_3=-9\mu H^2+3X(G_{2{X}}+2XG_{{2XX}})+$$$$+6X(3G^{(x)}_{XX}X\dot{\phi}H+6G^{(x)}_{X}\dot{\phi}H-G_{\phi})+$$
\begin{equation}+6H^2X(2G_{5XXX}X^2\dot{\phi}H+13G_{5XX}X\dot{\phi}H+15G_{5X}\dot{\phi}H).\label{w3}\end{equation}

 We see that the gravitational equations and the characteristics of perturbations ($c_T$, $c_S$, $Q_S$, $Q_T$) contain $H$, $\dot{\phi}$, $\phi$ and the Horndeski potentials. If we choose potentials $G_i(X,\phi)$ following the standard algorithm, the result is not predictable for the scale factor $a(t)$, the agreement with observational data, and the stability conditions. Here we will follow a different path, working within the framework of  "the inverse problem"\, or the  designer method. If we choose the form $c_T$, $c_S$, $Q_S$, $Q_T$, $H(\dot{\phi})$ or their combination as an ansatz, then (\ref{cS})-(\ref{QT}), (\ref{1gr00}), (\ref{2gr11}) form a system of differential equations on $G_i(X,\phi)$, $\phi(t)$, $a(t)$. The purpose of this procedure is to find subclasses of  HG  within which cosmological solutions exist without ghosts, Laplace instabilities and consistency with some observational data. We will consider the details of the method in the next section.

\section{Restoring of potentials $G_i(X,\phi)$}

 In this section we will select the initial ansatzes, consider in detail and transform the system of equalities (\ref{cS})-(\ref{QT}), (\ref{1gr00}), (\ref{2gr11}) for the  designer procedure.

As an ansatz we assume that $Q_T$, $c^2_T$, $Q_S$, $c^2_S$ have non-negative constant values. This ansatz allows us to exclude ghosts and Laplace instabilities
at the initial stage. To avoid cumbersome expressions, let us consider the consequences of these assumptions step by step.

Since $Q_T$, $c^2_T$ are constants, it follows from (\ref{csT}) and (\ref{QT}),  that $w_1$, $w_4$ are constants:
\begin{equation}\label{w1w4const} w_1=4Q_T=const>0, \,\, w_4=4Q_Tc^2_T=const\geq0,\end{equation}
then from equalities (\ref{w1w4}) it follows $\mu-2G_{5X}X\dot{\phi}H=4Q_T$, $\mu-2G_{5X}X\ddot{\phi}=4Q_Tc^2_T$, or
\begin{equation}\label{ddphi}\ddot{\phi}=-\beta H\dot{\phi},\end{equation}
\begin{equation}\label{q5x} G_{5X}=\frac{\mu-4Q_T}{2H\dot{\phi}X},\end{equation}
where
\begin{equation}\label{bett}\beta\equiv-\frac{c^2_T-\mu/(4Q_T)}{1-\mu/(4Q_T)}=-1+\frac{1-c^2_T}{1-\mu/(4Q_T)}.\end{equation}
For the speed $c_T$ we will use the estimate \cite{LIGO}:
\begin{equation}\label{malct}1-3\cdot 10^{-15}<c_T<1.\end{equation}
 The form of equations (\ref{ddphi}), (\ref{q5x}) leads us to the following reasoning. Since $G_5$ is a function of $X$, it is natural to choose  $H=H(\dot{\phi})$ dependence as the next ansatz. Then function $G_5(X)$ is restored from the formula (\ref{q5x}).  Function $H(\dot{\phi})$ and the equation (\ref{ddphi}) give us the following. Firstly, it is a dependency $\ddot{\phi}=-\beta H(\dot{\phi})\dot{\phi}=f(\dot{\phi})$. We can exclude $\ddot{\phi}$ from the system of remaining equations. In this case, the system coefficients will depend only on $X$ and $\phi$. In such system, potentials $G_i(X,\phi)$ may have the status of unknown functions. Secondly,  from system $H=H(\dot{\phi})$, (\ref{ddphi}) follow connections: $t(\dot{\phi})$, $H(t)$ and $a(t)$, $a[t(\dot{\phi})]$. These connections are used in the further procedure. Later, in other sections we will consider several cases of $H(\dot{\phi})$.

Let's return to the previous ansatz. Since $Q_S$, $c^2_S$ are constants, the combinations (\ref{cS}) and (\ref{QS}) give the following results:
\begin{equation}\label{ww3} w_3=\frac{3w^2_2}{4w_1}\left(\frac{Q_S}{w_1}-3\right)=\frac{3w^2_2}{16Q_T}\left(\frac{r}{16}-3\right),\end{equation}
\begin{equation}\label{bett37}\frac{d}{dt}\left(\frac{1}{w_2}\right)+\frac{H}{w_2}=A,\end{equation}
where
\begin{equation}\label{3odm}A\equiv\frac{Q_Sc^2_S+w_4}{2w_1^2}=\frac{1}{8Q_T}\left(c^2_T+c^2_S\cdot\frac{r}{16}\right)=\text{const}>0.\end{equation}
 Here we have taken into account that $w_1$, $w_4$ are constants (\ref{w1w4const}). Since $Q_T$, $Q_S$ are constants, the tensor-to-scalar ratio $r\equiv\dfrac{4Q_S}{Q_T}>0$ is a constant value. We believe according to \cite{Aghanim}:
\begin{equation}0<r<r_0=0.1.\end{equation}
 Equation (\ref{bett37}) has a particular solution:
\begin{equation}\label{ww2}w_2=\frac{a}{A\int adt}.\end{equation}
It is clear that the right side of equation (\ref{ww2}) depends on time $t$, and therefore in equation (\ref{ww3}) as well. Ansatz $H(\dot{\phi})$ defines relationship $t(\dot{\phi})$, therefore we obtain the form of functions $w_2=w_2(\dot{\phi})$, $w_3=w_3(\dot{\phi})$.  Substituting them into equations (\ref{w2}), (\ref{w3}), the next equations for the Horndeski potentials are derived.

From (\ref{w2}) it follows
\begin{equation}\label{gx} G^{(x)}_{X}=-\frac{w_2(\dot{\phi})}{2X\dot{\phi}}+\frac{H(\dot{\phi})}{2X\dot{\phi}}\left[20Q_T-3\mu+2(\mu-4Q_T)XL\right],\end{equation}
where $L\equiv (\ln|H\dot{\phi}X|)'_X$ and we used (\ref{q5x}). From here we can recover $G^{(x)}(X)$, which is part of function $G_3(X,\phi)$ (\ref{g3}).
From (\ref{w3}) it follows
$$XG_{2{X}}+2X^2G_{{2XX}}-2XG_{\phi}-\frac{w_3(\dot{\phi})}{3}-\frac{3Hw_2(\dot{\phi})}{2}-3HXw_{2X}'+$$
$$+H^2\Big\{6(10Q_T-\mu)+2(7\mu-40Q_T)XL-$$$$-4(\mu-4Q_T)X^2L^2+4(\mu-4Q_T)X^2L'_{X}+$$
\begin{equation}+3X(\ln H^2)'_X[20Q_T-3\mu+2(\mu-4Q_T)X L]\Big\},\label{rg1}\end{equation}
where we took into account (\ref{q5x}) and (\ref{gx}). Thus, the ansatz on the four constants ($c_T$, $c_S$, $Q_S$, $Q_T$) gave us four equalities (\ref{ddphi}), (\ref{q5x}), (\ref{gx}), (\ref{rg1}).

Taking into account equalities (\ref{ddphi}), (\ref{q5x}), (\ref{gx}) and the ansatz $H(\dot{\phi})$, from gravitational equations (\ref{1gr00}), (\ref{2gr11}) we obtain:
\begin{equation}-G_2+2XG_{2X}-2XG_\phi-3w_2(\dot{\phi})H+H^2[40Q_T-7\mu+4(\mu-4Q_T)XL]=0,\label{rg2}\end{equation}
$$-G_2+2XG_\phi+\beta w_2(\dot{\phi}) H-H^2\Big\{\mu+8Q_T+\beta(20Q_T-3\mu)+2\beta(\mu-4Q_T)XL-$$
\begin{equation}-\beta (\mu+4Q_T)X (\ln H^2)'_X\Big\}=0.\label{rg3}\end{equation}
Equations (\ref{rg1}), (\ref{rg2}), (\ref{rg3}) form a system of differential equations for functions $G_2(X,\phi)$, $G_{\phi}$. The system is overdetermined -- there are two unknown functions and three equations. However, the equations contain partial derivatives of $G_2(X,\phi)$. Consequently, during integration, arbitrary functions arise, which allows satisfying the “excess”\, equation.

 Given a known ansatz $H(\dot{\phi})$, the system (\ref{q5x}), (\ref{gx})-(\ref{rg3})  will determine all Horndeski potentials. In the following sections, we will consider the {\bf designer}  results for different types of $H(\dot{\phi})$.

\section{De Sitter world}

 Exponential expansion helped solve cosmological problems: isotropy, homogeneity, flatness,
the horizon and some other problems \cite{Guth,Linde}.  In article \cite{Sato}, the first-order phase transition model of the early Universe that leads to an exponential expansion which stretches domains much greater than the horizon scales was studied.  Fluctuations associated with the phase transition are exponentially stretched and then may play the role of seed fluctuations for large-scale structures \cite{Sato1}. The inflationary Universe is one of the central objects of study in many investigations \cite{Belinsky, Piran, Piran1, Halliwell}.

Here we choose the ansatz $H=\text{const}>0$. In this case, the space-time expands exponentially (de Sitter world):
\begin{equation}a=a_0e^{Ht},\, t\in (-\infty,+\infty).\end{equation}
The scalar field and its derivative have the following time dependence (see (\ref{ddphi})):
\begin{equation}\label{qwe}\dot{\phi}=\frac{\phi_0e^{-\beta Ht}}{a_0^\beta},\,\, \phi=-\frac{\phi_0e^{-\beta Ht}}{\beta H a_0^\beta}.\end{equation}
In case $\beta>0$, the scalar field is singular since one has $|\phi|, |\dot{\phi}|\rightarrow \infty$ as $t\rightarrow -\infty$.

 For the chosen ansatz $H$, auxiliary functions $L$ and $w_2$, $w_3$ (see (\ref{ww3}), (\ref{ww2})) are equal $L\equiv(\ln|H\dot{\phi}X|)'_X=3/(2X)$,
$w_2=H/A$, $w_3=\dfrac{3H^2}{16A^2Q_T}\left(r/16-3\right)$.
Then the system (\ref{q5x}), (\ref{gx})-(\ref{rg3}) defining the Horndeski potentials takes the form:
\begin{equation}\label{3mdkg}G_{5X}=\epsilon\cdot\frac{\mu-4Q_T}{2\sqrt{2} H X^{3/2}},\end{equation}
\begin{equation}\label{3m8g}G^{(x)}_{X}=\frac{\epsilon H}{2\sqrt{2}}(-1/A+8Q_T)X^{-3/2},\end{equation}
\begin{equation}\label{3q0m8g}XG_{2{X}}+2X^2G_{{2XX}}-2XG_{\phi}-H^2\left[\frac{3}{2A}+\frac{r/16-3}{16A^2Q_T}\right]=0, \end{equation}
\begin{equation}\label{3m8zmg} -G_2+2XG_{2X}-2XG_\phi+H^2\left[-3/A-\mu+16Q_T\right]=0,\end{equation}
\begin{equation}\label{2pm8g}-G_2+2XG_\phi+H^2[\beta/A-\mu-(\beta+1)8Q_T]=0, \end{equation}
where $\epsilon=\pm 1$  is  the sign of the derivative $\dot{\phi}$ (or $\phi_0$ in (\ref{qwe})).

  Potentials $G_5(X)$, $G^{(x)}(X)$ are easily determined from equations (\ref{3mdkg}), (\ref{3m8g}):
\begin{equation}\label{do2b} G_5=-\frac{\mu-4Q_T}{\sqrt{2} H}\cdot \epsilon X^{-1/2}+\text{const},\end{equation}
\begin{equation} G^{(x)}=-\frac{\epsilon H}{\sqrt{2}}(-1/A+8Q_T)X^{-1/2}+\text{const}.\end{equation}

The subsystem (\ref{3q0m8g})-(\ref{2pm8g}) has a solution
\begin{equation}\label{d8s} G_2=S(\phi)X-H^2\left(\frac{3-\beta}{2A}+\mu+(\beta-1)4Q_T\right),\end{equation}
\begin{equation}G=\frac{1}{2}\int S(\phi)d\phi-\frac{(\beta+3)\phi^{-1}}{2\beta^2}(-1/A+8Q_T),\end{equation}
where $S(\phi)$ -- the arbitrary function.
From (\ref{g3}) we find function $G_3 (X,\phi)$:
$$G_3=-\frac{\epsilon H}{\sqrt{2}}(-1/A+8Q_T)X^{-1/2}+\text{const}+$$
\begin{equation}+\frac{1}{2}\int S(\phi)d\phi-\frac{(\beta+3)\phi^{-1}}{2\beta^2}(-1/A+8Q_T).\label{e4kw}\end{equation}
Thus, we have restored all of the Horndeski potentials.

 Since the subsystem (\ref{3q0m8g})-(\ref{2pm8g}) is redefined, a condition arises on the parameters of the found subclass HG:
\begin{equation}\label{conpar1}(3+\beta)(8AQ_T)(1-(8AQ_T))=3(r/48-1+(8AQ_T)).\end{equation}
Next, we need to make sure that equality (\ref{conpar1}) does not contradict the conditions of the absence of ghosts and Laplace instability.
Knowing $AQ_T$ from the formula (\ref{3odm}), we obtain the parameter $\beta$ from the last equality:
$$\beta=\frac{3[r/48+(8AQ_T)^2-1]}{8AQ_T(1-8AQ_T)}=$$
$$=\frac{3}{(c_T^2+c^2_S
\cdot\frac{r}{16})\left(1-c_T^2-c^2_S\cdot\frac{r}{16}\right)}\times$$
\begin{equation}\times\left[r/48+\left(c_T^2+c^2_S
\cdot\frac{r}{16}\right)^2-1\right].\label{s7gh}\end{equation}
This equality, in combination with the definition (\ref{bett}), gives the equality
\begin{equation}1-\frac{\mu}{4Q_T}=\frac{(1-c_T^2)\left(1-c_T^2-c^2_S\cdot\frac{r}{16}\right)(c_T^2+c^2_S
\cdot\frac{r}{16})}{r/16+\left(c_T^2-1+c^2_S\cdot\frac{r}{16}\right)\left(3+2c_T^2+2c^2_S
\cdot\frac{r}{16}\right)}.\label{gfhh}\end{equation}
 The score of $\beta$ and $\frac{Q_T}{\mu}$ depend on the estimates of $c^2_S\cdot\frac{r}{16}$ relative to $1-c^2_T$. We will consider two cases.

1. In case $c^2_S\cdot\frac{r}{16}\gg 1-c^2_T>0$, we get the approximation
\begin{equation}\frac{Q_T}{\mu}\approx \frac{1}{4}-\frac{1}{4}\cdot\frac{c_S^2\left(1+c^2_S
\cdot\frac{r}{16}\right)}{1+c_S^2\left(5+2c_S^2\cdot\frac{r}{16}\right)}(1-c_T^2)>0,\label{doli}\end{equation}
i.e. there are no pathologies:  $Q_T\approx \mu/4>0$, $Q_S=rQ_T/4>0$.
The parameter $\beta$ has a negative value:
\begin{equation}\label{b2jw}\beta\approx-\frac{1+c_S^2\left(6+3c^2_S \cdot\frac{r}{16}\right)}{c_S^2\left(1+c^2_S
\cdot\frac{r}{16}\right)}<0,\end{equation}
then the scalar field does not have the initial singularity (see (\ref{qwe})): $\phi\rightarrow 0$ as $t\rightarrow -\infty$. The value of $|\phi(t)|$ increases over time.

2. In case $c^2_S\cdot\frac{r}{16}=k^2(1-c_T^2)\ll 1$ ($k\neq 1$) from equality (\ref{s7gh}) and (\ref{gfhh}) it follows
$$\beta=\frac{3}{(1-k^2)(1-c_T^2)(1+(k^2-1)(1-c_T^2))}\times$$
\begin{equation}\times\left[r/48+2(1-k^2)(1-c_T^2)+(k^2-1)^2(1-c_T^2)^2\right],\end{equation}
\begin{equation}1-\frac{\mu}{4Q_T}=\frac{(1-k^2)(1-c_T^2)^2(1+(k^2-1)(1-c_T^2))}{r/16+(k^2-1)(1-c_T^2)(5+2(k^2-1)(1-c_T^2))}.\end{equation}
Assuming $|1-k^2|(1-c_T^2)\ll r$ and (\ref{malct}), we get
\begin{equation}\frac{Q_T}{\mu}\approx \frac{1}{4}+\frac{4}{r}(1-k^2)(1-c_T^2)^2>0,\label{doem}\end{equation}
\begin{equation}\label{doem1}\beta\approx\frac{r}{16(1-k^2)(1-c_T^2)}.\end{equation}
Inequalities $Q_T>0$, $Q_S>0$ remain valid. As we see $|\beta|\gg 1$. For $k>1$, the value $\beta$ is negative, then the scalar field does not have the initial singularity. The value of $|\phi(t)|$ increases over time.
For $k<1$, the value of $|\phi(t)|$ decreases over time. Thus, equality (\ref{conpar1}) does not contradict the conditions of the absence of ghosts and Laplace instability.

\section{The power-law inflation of the Universe}

The exponential growth of the scale factor is just a particular case of the inflationary expansion.
Power-law inflation also has some interest among researchers \cite{Abbott,Lucchin,Barrow1, Muller}. Detailed investigations of power-law inflation have been carried out in the work \cite{Lucchin}. The authors obtained constraints on this model coming from the requirement of solving the horizon, flatness, "good"\, reheating, and "convenient"\, perturbation spectrum problems. An exact
power-law inflationary solution possessing an exponential potential was given in the work \cite{Barrow1}.

Here we will assume the ansatz $H\sim\dot{\phi}^m$. In this case, we obtain power-law of the Universe expansion, $a\sim t^n$. In addition to the main task, we will find conditions on the parameters $n$ that ensure the accelerated expansion regime.

\subsection{Case $H=\gamma \dot{\phi}$}

Here we choose the ansatz $H=\gamma \dot{\phi}$. From equation (\ref{ddphi}) we find the Hubble parameter and the scalar factor:
\begin{equation}H=\frac{1}{\beta t},\end{equation}
\begin{equation}\label{fact1} a=a_0(t/t_0)^{1/\beta},\,\, t\geq 0, t_0>0; a\in (0,+\infty).\end{equation}
If $\beta>0$, then the model describes the expanding Universe. If $0<\beta<1$, then there is accelerated expansion.
For $\beta<0$ there is the solution with the big rip:
\begin{equation}a=a_0\left(\frac{t_0}{-t}\right)^{1/|\beta|},\,\, t\leq 0, t_0>0; a\in (0,+\infty).\end{equation}
We will consider the model (\ref{fact1}) with $\beta>0$.

The scalar field has a dependence
\begin{equation}\phi=\frac{1}{\gamma}\ln(a/c_1).\end{equation}
The scalar field is singular since one has $|\phi|\rightarrow \infty$ as $a\rightarrow 0$. The value of $|\phi(t)|$ increases over time.
Next we may need equalities:
\begin{equation}\dot{\phi}=\frac{1}{\beta\gamma t}=\frac{(a_0/c_1)^\beta}{\beta\gamma t_0}\cdot e^{-\gamma\beta\phi}.\end{equation}

 For the chosen ansatz $H=\gamma\dot{\phi}$, auxiliary functions $L$ and $w_2$, $w_3$ (see (\ref{ww3}), (\ref{ww2})) are equal
$$L=2/X,\,\, w_2=\frac{\beta+1}{A\beta t }=\frac{(\beta+1)\gamma \dot{\phi}}{A},$$
\begin{equation}w_3=X\cdot\frac{6\gamma^2(\beta+1)^2}{16Q_TA^2}\left(r/16-3\right).\end{equation}
Then the system (\ref{q5x}), (\ref{gx})-(\ref{rg3}) defining the Horndeski potentials takes the form:
\begin{equation}\label{sd12} G_{5X}=\frac{\mu-4Q_T}{4\gamma X^{2}},\end{equation}
\begin{equation}\label{sd13} G^{(x)}_{X}=\frac{\gamma}{2X}\left[-\frac{\beta+1}{A}+\mu+4Q_T\right],\end{equation}
\begin{equation}\label{sd14}G_{2{X}}+2XG_{{2XX}}-2G_{\phi}+2\gamma^2\Big\{\frac{(3-r/16)(\beta+1)^2}{16A^2Q_T}-\frac{3(\beta+1)}{A}+\mu+8Q_T\Big\}=0,\end{equation}
\begin{equation}\label{sd15}-G_2+2XG_{2X}-2XG_\phi+2\gamma^2X\Big\{-\frac{3(\beta+1)}{A}+\mu+8Q_T\Big\}=0,\end{equation}
\begin{equation}\label{sd16}-G_2+2XG_\phi+2\gamma^2X\Big\{\frac{\beta(\beta+1)}{A}-\mu-8Q_T\Big\}=0.\end{equation}

  Potentials $G_5(X)$, $G^{(x)}(X)$ are easily determined from equations (\ref{sd12}), (\ref{sd13}):
\begin{equation}G_5=-\frac{\mu-4Q_T}{4\gamma}\cdot X^{-1}+\text{const}.\end{equation}
\begin{equation} G^{(x)}=\frac{\gamma}{2}\left[-\frac{\beta+1}{A}+\mu+4Q_T\right]\cdot\ln X+\text{const}.\end{equation}

The subsystem (\ref{sd14})-(\ref{sd16}) has a solution:
\begin{equation}G_2=\varepsilon X-V,\,\, V=\frac{(3-\beta)(\beta+1)(a_0/c_1)^{2\beta}}{2 A(\beta t_0)^2}\cdot e^{-2\gamma\beta\phi},\end{equation}
\begin{equation}G=\phi\cdot\Big\{\varepsilon/2+\gamma^2\Big(-\frac{(\beta+3)(\beta+1)}{2A}+\mu+8Q_T\Big)\Big\}.\end{equation}
From (\ref{g3}) we find function $G_3 (X,\phi)$:
$$G_3=\frac{\gamma}{2}\left[-\frac{\beta+1}{A}+\mu+4Q_T\right]\cdot\ln X+\text{const}+$$
\begin{equation}+\phi\cdot\Big\{\varepsilon/2+\gamma^2\Big(-\frac{(\beta+3)(\beta+1)}{2A}+\mu+8Q_T\Big)\Big\}.\end{equation}
Thus, we have restored all of the Horndeski potentials.

 Since the subsystem (\ref{sd14})-(\ref{sd16}) is redefined, a condition arises on the parameters of the found subclass HG:
\begin{equation}\label{so1d}\frac{\beta-3}{2 A}+\frac{(\beta+1)(3-r/16)}{16A^2Q_T}=0.\end{equation}
Next, we need to make sure that equality (\ref{so1d}) does not contradict the conditions of the absence of ghosts and Laplace instability. We will also check the fulfillment of the condition $0<\beta<1$ of accelerated expansion of space-time
Knowing $AQ_T$ from the formula (\ref{3odm}), we obtain the parameter $\beta$ from the last equality:
\begin{equation}\label{sla}\beta=\frac{3(8AQ_T-1)+r/16}{3+8AQ_T-r/16}=\frac{3(c^2_T-1)+(3c^2_S+1)r/16}{3+c^2_T+(c^2_S-1)r/16}.\end{equation}
The condition of accelerated expansion is satisfied when
\begin{equation}\label{neravcs}(1-c^2_T)16/r-1/3<c^2_S<(3-c^2_T)16/r-1.\end{equation}
From (\ref{bett}) and (\ref{sla}) it follows
\begin{equation}\label{seij}\frac{4Q_T}{\mu}=\left[1-\frac{3+c^2_T+(c^2_S-1)r/16}{4(c^2_T+c^2_Sr/16)}\cdot(1-c^2_T)\right]^{-1}.\end{equation}
 The score of $\beta$ and $\frac{Q_T}{\mu}$ depend on the estimates of $c^2_S\cdot\frac{r}{16}$ relative to $1-c^2_T$. We will consider two cases.

1. In case $c^2_S\cdot\frac{r}{16}\gg 1-c^2_T$, we get the approximation
\begin{equation}\frac{Q_T}{\mu}\approx \frac{1}{4}+\frac{1+(c^2_S-1)r/64}{4(1+c^2_Sr/16)}\cdot(1-c^2_T),\end{equation}
\begin{equation}\beta\approx\frac{(3c^2_S+1)r/16}{4+(c^2_S-1)r/16}.\end{equation}
Condition of accelerated mode (\ref{neravcs}) is rewritten as follows
\begin{equation}-1/3<c^2_S<32/r-1.\end{equation}
It is fulfilled because $c^2_S\geq0$, $r<r_0=0.1$.  The conditions for the absence of ghosts are met:  $Q_T\approx \mu/4>0$, $Q_S=rQ_T/4>0$.

2. In case $c^2_S\cdot\frac{r}{16}=k^2(1-c_T^2)\ll 1$ ($k\neq 1$) from equalities (\ref{sla}) and (\ref{seij}) it follows
$$\beta=\frac{r/16+3(k^2-1)(1-c^2_T)}{4-r/16+(k^2-1)(1-c^2_T)}\approx$$\begin{equation}\approx \frac{r/16}{4-r/16}<r/16<0.00625,\end{equation}
$$\frac{Q_T}{\mu}=\frac{1}{4}\left[1-\frac{4-r/16+(k^2-1)(1-c^2_T)}{4(1+(k^2-1)(1-c^2_T))}\cdot(1-c^2_T)\right]^{-1}\approx$$
\begin{equation}\approx\frac{1}{4}+\frac{1}{4}(1-r/64)(1-c^2_T).\end{equation}
The condition of accelerated expansion $0<\beta<1$ is satisfied. The conditions for the absence of ghosts are confirmed: $Q_T\approx \mu/4>0$, $Q_S=rQ_T/4>0$.

\subsection{Case $H=\gamma_1 |\dot{\phi}|^{1/2}$, $\gamma_1>0$}

The Hubble parameter and the scale factor are
\begin{equation}H=\frac{2}{\beta t},\end{equation}
\begin{equation}\label{fact2}a=a_0(t/t_0)^{2/\beta},\,\, t\geq 0, t_0>0; a\in (0,+\infty).\end{equation}
We will assume $\beta>0$. If $0<\beta<2$, then there is accelerated expansion of the Universe.

The scalar field has a dependence
\begin{equation}\phi=-\frac{4\epsilon}{\gamma_1^2\beta^2 t},\end{equation}
where $\epsilon=\pm 1$  is  the sign of the derivative $\dot{\phi}$:
\begin{equation}\dot{\phi}=\frac{4\epsilon}{\gamma_1^2\beta^2 t^2}=\frac{\epsilon\gamma_1^2\beta^2}{4}\cdot\phi^2.\end{equation}
The scalar field is singular since one has $|\phi|\rightarrow \infty$ as $t\rightarrow 0$. The value of $|\phi(t)|$ decreases over time.

Reasoning in the same way as in the previous section, we get $G_5(X)$, $G^{(x)}(X)$:
\begin{equation}G_5=-\frac{2^{1/4}\epsilon}{3}\cdot\frac{\mu-4Q_T}{\gamma_1}\cdot X^{-3/4}+\text{const},\end{equation}
\begin{equation}G^{(x)}=-\frac{\epsilon\gamma_1}{(2X)^{1/4}}\left[-\frac{\beta+2}{A}+\mu+12Q_T\right]+\text{const}.\end{equation}

The system (\ref{rg1}), (\ref{rg2}), (\ref{rg3}) will change as follows:
$$XG_{2{X}}+2X^2G_{{2XX}}-2XG_{\phi}+$$\begin{equation}+2^{1/2}X^{1/2}\gamma_1^2
\Big\{\frac{(3-r/16)(\beta+2)^2}{64A^2Q_T}-\frac{9(\beta+2)}{8A}+6Q_T\Big\}=0,\label{rg11}\end{equation}
$$-G_2+2XG_{2X}-2XG_\phi+$$\begin{equation}+2^{1/2}X^{1/2}\gamma_1^2
\Big\{-\frac{3(\beta+2)}{2A}+12Q_T\Big\}=0,\label{rg22}\end{equation}
\begin{equation}-G_2+2XG_\phi+2^{1/2}X^{1/2}\gamma_1^2
\Big\{\frac{\beta(\beta+2)}{2A}-\mu-4(\beta+2)Q_T\Big\}=0.\label{rg33}\end{equation}
 We will consider two solutions for this system.

The solution I:
$$G_2=\varepsilon X-V,$$\begin{equation}V=\phi^2\cdot\frac{\gamma_1^4\beta^2}{16}\Big\{\frac{(3-\beta)(\beta+2)}{A}+2(\mu+4(\beta-1)Q_T)\Big\},\end{equation}
\begin{equation}G=\frac{\varepsilon}{2}\cdot\phi-\beta^{-2}\phi^{-1}\Big\{-\frac{(\beta+3)(\beta+2)}{A}+2\mu+8(5+\beta)Q_T\Big\}.\end{equation}
From (\ref{g3}) we find function $G_3 (X,\phi)$:
$$G_3=-\frac{\epsilon\gamma_1}{(2X)^{1/4}}\left[-\frac{\beta+2}{A}+\mu+12Q_T\right]+\text{const}+$$
\begin{equation}+\frac{\varepsilon}{2}\cdot\phi-\beta^{-2}\phi^{-1}\Big\{-\frac{(\beta+3)(\beta+2)}{A}+2\mu+8(5+\beta)Q_T\Big\}.\end{equation}
The system (\ref{rg11})-(\ref{rg33}) also provides restrictions on parameters:
\begin{equation}\label{so1d2}\frac{(\beta+2)(2\beta-3)}{8A}-\mu/2-2(\beta+2)Q_T+\frac{(3-r/16)(\beta+2)^2}{64A^2Q_T}=0.\end{equation}
According to (\ref{3odm}), the parameter $AQ_T$ depends on $c_T$, $c_S$ and $r$.
 Next, we need to make sure that equality (\ref{so1d2}) does not contradict the conditions of the absence of ghosts and Laplace instability. We will also check the fulfillment of the condition $0<\beta<2$ of accelerated expansion of space-time.
In case $1-c^2_T\ll r/16\ll 1$ and $1-c^2_T\ll c^2_Sr/16\ll 1$, we get one of the approximations
\begin{equation}\beta\approx\frac{2(2+9c^2_S)}{11}\cdot r/16>0.\end{equation}
 This equality, in combination with the definition (\ref{bett}), gives the equality
\begin{equation}\frac{4Q_T}{\mu}\approx 1+(1-c^2_T)\left(1-\frac{2(2+9c^2_S)}{11}\cdot r/16\right).\end{equation}
 The parameter $0<\beta\ll 1$, that is, the condition of inflationary expansion is met. The conditions for the absence of ghosts are confirmed: $Q_T\approx \mu/4>0$, $Q_S=rQ_T/4>0$.

The solution II:
\begin{equation}G_2=S(\phi)X+X^{1/2}\cdot 2^{-1/2}\gamma_1^2\left[\frac{(\beta-3)(\beta+2)}{A}-2(\mu+4(\beta-1)Q_T)\right],\end{equation}
\begin{equation}G=\frac{1}{2}\int S(\phi)d\phi-\frac{2\phi^{-1}}{\beta^2}\left[-\frac{3(\beta+2)}{A}+24Q_T\right],\end{equation}
where $S(\phi)$ -- the arbitrary function.
From (\ref{g3}) we find function $G_3 (X,\phi)$:
$$G_3=-\frac{\epsilon\gamma_1}{(2X)^{1/4}}\left[-\frac{\beta+2}{A}+\mu+12Q_T\right]+\text{const}+$$
\begin{equation}+\frac{1}{2}\int S(\phi)d\phi-\frac{2\phi^{-1}}{\beta^2}\left[-\frac{3(\beta+2)}{A}+24Q_T\right].\end{equation}

 The system also provides restrictions on parameters:
\begin{equation}(1-r/48)\left(\frac{\beta+2}{8AQ_T}\right)^2+\frac{\beta+2}{8AQ_T}-2=0.\end{equation}
The equation has roots
\begin{equation}\beta=-2+\frac{8AQ_T}{2(1-r/48)}\cdot\left[\pm3\sqrt{1-r/54}-1\right].\end{equation}

In case $r/48\ll 1$, $c^2_Sr/16\gg 1-c^2_T$, the positive root has an approximation
\begin{equation}\beta\approx-1+c^2_Sr/16,\end{equation}
and
\begin{equation}\frac{Q_T}{\mu}\approx 1/4+(1-c^2_T)\cdot\frac{4}{c^2_Sr}.\end{equation}
The condition of accelerated expansion $0<\beta<2$ is satisfied, when $1<c^2_Sr/16<3$. The conditions for the absence of ghosts are confirmed: $Q_T\approx \mu/4>0$, $Q_S=rQ_T/4>0$.

\subsection{General case $H=\gamma_n \dot{\phi}^{2 n}$}

Here we will assume that $H=\gamma_n \dot{\phi}^{2 n}$, $\gamma_n>0$, $\dot{\phi}>0$. Special cases $n=0$ ($H$=const), $n=1/2$ ($H=\gamma \dot{\phi}$) were discussed above. The Hubble parameter and the scale factor are
\begin{equation}H=\frac{1}{2n\beta t},\,\, n\neq0,\end{equation}
\begin{equation}\label{fact2}a=a_0(t/t_0)^{1/(2n\beta)},\,\, t\geq 0, \,\,t_0>0; a\in (0,+\infty).\end{equation}
We will assume $2n\beta>0$. If $0<2n\beta<1$, then there is accelerated expansion of the Universe.

The scalar field and its derivative have the following time dependence:
\begin{equation}\dot{\phi}^{2 n}=\frac{1}{2n\gamma_n\beta t}=[(2n-1)\gamma_n\beta\phi]^{-\frac{2n}{2n-1}},\,\,\phi=\frac{2n\cdot t^{1-1/(2n)}}{(2n-1)(2n\gamma_n\beta)^{1/(2n)}},\,\, n\neq 1/2.\end{equation}
The case $n=1/2$ ($H=\gamma \dot{\phi}$) was discussed above.
For $0<n<1/2$ the scalar field is singular since one has $|\phi|\rightarrow \infty$ as $t\rightarrow 0$ and the value of $|\phi(t)|$ decreases over time. In the case of the
 arbitrary degree $1-1/(2n)$, for the signs of $t$ and $\phi$ to correspond, we assume the condition $2n/(2n-1)>0$, therefore $n\in (-\infty,0)\cup (1/2,+\infty)$.

 For the chosen ansatz $H=\gamma_n \dot{\phi}^{2 n}$, auxiliary functions $L$ and $w_2$, $w_3$ (see (\ref{ww3}), (\ref{ww2})) are equal
$$L=(3/2+n)/X,\,\, w_2=\frac{2n\beta+1}{2 A n\beta t }=\frac{(2n\beta+1)\gamma_n \dot{\phi}^{2 n}}{A},$$
\begin{equation}w_3=\frac{3\gamma_n^2\dot{\phi}^{4 n}(2n\beta+1)^2}{16A^2Q_T}\left(r/16-3\right).\end{equation}
Then the system (\ref{q5x}), (\ref{gx})-(\ref{rg3}) defining the Horndeski potentials takes the form:
\begin{equation}\label{sd12w} G_{5X}=\frac{\mu-4Q_T}{\gamma_n(2X)^{n+3/2}}, \end{equation}
\begin{equation}\label{sd13w} G^{(x)}_{X}=X^{n-3/2}\cdot2^{n-3/2}\gamma_n\left[-\frac{2n\beta+1}{A}+2n\mu+8(1-n)Q_T\right],\end{equation}
$$XG_{2{X}}+2X^2G_{{2XX}}-2XG_{\phi}+2^{2n}\gamma_n^2X^{2n}
\Big\{\frac{(3-r/16)(2n\beta+1)^2}{16A^2Q_T}-$$\begin{equation}-\frac{3(2n+1)(2n\beta+1)}{2A}+2n(4n-1)\mu+32n(1-n)Q_T\Big\}=0,\label{syk}\end{equation}
$$-G_2+2XG_{2X}-2XG_\phi+$$\begin{equation}+2^{2n}\gamma_n^2X^{2n}
\Big\{-\frac{3(2n\beta+1)}{A}+(4n-1)\mu+16(1-n)Q_T\Big\}=0,\label{syk1}\end{equation}
\begin{equation}-G_2+2XG_\phi+2^{2n}\gamma_n^2X^{2n}
\Big\{\frac{\beta(2n\beta+1)}{A}-\mu-8(1+\beta(1-2n))Q_T\Big\}=0.\label{syk2}\end{equation}

  Potentials $G_5(X)$, $G^{(x)}(X)$ are determined from equations (\ref{sd12w}), (\ref{sd13w}):
\begin{equation} G_5=-\frac{2^{-n-1/2}}{2n+1}\cdot\frac{\mu-4Q_T}{\gamma_n}\cdot X^{-n-1/2}+\text{const},\end{equation}
\begin{equation}G^{(x)}=X^{n-1/2}\cdot\frac{2^{n-1/2}\gamma_n}{2n-1}\left[-\frac{2n\beta+1}{A}+2n\mu+8(1-n)Q_T\right].\end{equation}
 We will consider two solutions for the subsystem (\ref{syk})-(\ref{syk2}).

The solution I:
$$G_2=\varepsilon X-V,$$
\begin{equation}V=\frac{\phi^{-\frac{4n}{2n-1}}\gamma_n^{-\frac{2}{2n-1}}}{[(2n-1)\beta]^{\frac{4n}{2n-1}}}\Big\{\frac{(3-\beta)(2n\beta+1)}{2A}+(1-2n)(\mu+4(\beta-1)Q_T)\Big\},\end{equation}
\begin{equation}G=\frac{\varepsilon}{2}\cdot\phi-\frac{\phi^{-1}}{(2n-1)^{2}\beta^{2}}\Big\{-\frac{(\beta+3)(2n\beta+1)}{2A}+2n\mu+4(3-2n+(1-2n)\beta)Q_T\Big\}.\end{equation}
From (\ref{g3}) we find function $G_3 (X,\phi)$:
$$G_3=X^{n-1/2}\cdot\frac{2^{n-1/2}\gamma_n}{2n-1}\left[-\frac{2n\beta+1}{A}+2n\mu+8(1-n)Q_T\right]+$$
\begin{equation}+\frac{\varepsilon}{2}\cdot\phi-\frac{\phi^{-1}}{(2n-1)^{2}\beta^{2}}\Big\{-\frac{(\beta+3)(2n\beta+1)}{2A}+2n\mu+4(3-2n+(1-2n)\beta)Q_T\Big\}.\end{equation}
 The system also provides restrictions on parameters:
$$\frac{(2n\beta+1)(\beta-6n)}{2A}+4n(2n-1)\mu+$$\begin{equation}+4(2n-1)(\beta-4n+3)Q_T+\frac{(3-r/16)(2n\beta+1)^2}{16A^2Q_T}=0.\label{3dn}\end{equation}
According to (\ref{3odm}), the parameter $AQ_T$ depends on $c_T$, $c_S$ and $r$.

In case $1-c^2_T\ll r/16\ll 1$ and $1-c^2_T\ll c^2_Sr/16\ll 1$, we get one of the approximations
\begin{equation}\beta\approx \frac{1+6c^2_S(1-n)}{2n(7-6n)}\cdot\frac{r}{16}.\end{equation}
This equality, in combination with the definition (\ref{bett}), gives the equality
\begin{equation}\label{x7b}\frac{Q_T}{\mu}\approx \frac{1}{4}+\frac{1-c^2_T}{4}\cdot\Big[1-\frac{1+6c^2_S(1-n)}{2n(7-6n)}\cdot\frac{r}{16}\Big].\end{equation}
The condition of accelerated expansion $0<2n\beta<1$ is satisfied, when
\begin{equation}n\in(-\infty,7/6)\cup (1+1/(6c^2_S), +\infty),\end{equation}
where $0<c^2_S<1$. The value $n$ is estimated with an error of at least $r/16$.  From (\ref{x7b}) it is clear that there are many values of $n$ for which the conditions
for the absence of ghosts are confirmed.
Another solution to equations (\ref{3dn}) is
$$\beta\approx \frac{6n-7}{6n+1}+\frac{(2n-1)}{2n(6n+1)^2(6n-7)}\times$$\begin{equation}\times\{(2n-1)(6n-1)^2+c^2_S[1+6n(36n^2-54n+7)]\}\cdot\frac{r}{16},\end{equation}
 This equality, in combination with the definition (\ref{bett}), gives the equality
$$\frac{Q_T}{\mu}\approx \frac{1}{4}+\frac{1-c^2_T}{4}\cdot\Big[1-\frac{r}{16}\cdot\frac{1}{12n(6n+1)(6n-7)}\times$$
\begin{equation}\times\{(2n-1)(6n-1)^2+c^2_S[1+6n(36n^2-54n+7)]\}\Big].\label{qm8z}\end{equation}
The condition of accelerated expansion $0<2n\beta<1$ is satisfied, when
\begin{equation}\frac{7}{6}<n<\frac{5+2\sqrt{7}}{6},\,\, \text{where} \,\,\frac{7}{6}\approx 1.167,\,\, \frac{5+2\sqrt{7}}{6}\approx 1.715.\end{equation}
The value $n$ is estimated with an error of at least $r/16$.  As can be seen from (\ref{qm8z}), the conditions for the absence of ghosts are confirmed: $Q_T\approx \mu/4>0$, $Q_S=rQ_T/4>0$.

The solution II:
$$G_2=S(\phi)X-\frac{2^{2n}\gamma_n^2X^{2n}}{2n-1}\Big\{\frac{(\beta-3)(2n\beta+1)}{2A}+(2n-1)(\mu+4(\beta-1)Q_T)\Big\},$$
\begin{equation}G=\frac{1}{2}\int S(\phi)d\phi-\frac{\phi^{-1}}{(2n-1)^2\beta^2}\left\{\frac{2n\beta+1}{A}\left(\frac{3-\beta}{2(2n-1)}-\beta\right)+
4(3+\beta(1-4n))Q_T\right\},\end{equation}
where $S(\phi)$ -- the arbitrary function. From (\ref{g3}) we find function $G_3 (X,\phi)$:
$$G_3=X^{n-1/2}\cdot\frac{2^{n-1/2}\gamma_n}{2n-1}\left[-\frac{2n\beta+1}{A}+2n\mu+8(1-n)Q_T\right]+$$
\begin{equation}+\frac{1}{2}\int S(\phi)d\phi-\frac{\phi^{-1}}{(2n-1)^2\beta^2}\left\{\frac{2n\beta+1}{A}\left(\frac{3-\beta}{2(2n-1)}-\beta\right)+
4(3+\beta(1-4n))Q_T\right\}.\end{equation}
 The system also provides restrictions on parameters:
\begin{equation}\frac{2n\beta+1}{A}[(4n-1)\beta/2-3n]+4(2n-1)[(4n-1)\beta-3]Q_T-\frac{(3-r/16)(2n\beta+1)^2}{16A^2Q_T}=0.\end{equation}
The model was previously investigated in detail for case  $n=1/4$ ($H=\gamma_1 |\dot{\phi}|^{1/2}$).

\section{Estimates of Horndeski's potential in de Sitter world}

In the built models there is a small parameter $1-c^2_T$. In this section, using this parameter, we will estimate the contributions of $\mathcal{L}_i$ ($i=2,3,5$) relative to the Einstein-Hilbert term
$\mathcal{L}_4=\mu R/2$ in the action density (\ref{lagr1}). The estimation will be made in  de Sitter world.

We will need factors in the action density (\ref{lagr1}):
$$\Box \phi=-(\ddot{\phi}+3H\dot{\phi}), \, \, G_{\mu\nu}\,\nabla^\mu \nabla^\nu\phi=3H^2(\ddot{\phi}+3H\dot{\phi}),$$
\begin{equation}\left( \Box \phi \right)^3 -3 \Box\phi (\nabla_\mu \nabla_\nu \phi)^2 + 2\left(\nabla_\mu \nabla_\nu\phi \right)^3=
-6H^2\dot{\phi}^2(3\ddot{\phi}+H\dot{\phi}),\end{equation}
where $H$=const. Then, taking into account (\ref{ddphi}), we get
\begin{equation}\mathcal{L}_3 = (3-\beta)H\dot{\phi}G_3, \, \, \mathcal{L}_5=H^3\dot{\phi}[3G_5(3-\beta)+G_{5X}\dot{\phi}^2(1-3\beta)].\end{equation}
The Einstein-Hilbert term has the form
\begin{equation} \mathcal{L}_4=6\mu H^2.\end{equation}
For de Sitter world, we have restored the potentials of Horndeski (\ref{do2b}), (\ref{d8s}), (\ref{e4kw}).
Let's take special case $S(\phi)=0$:
$$G_2=-H^2\left(\frac{3-\beta}{2A}+\mu+(\beta-1)4Q_T\right),$$
\begin{equation}G_3=-(-1/A+8Q_T)\left(\frac{\epsilon H}{\sqrt{2}X^{1/2}}+\frac{(\beta+3)\phi^{-1}}{2\beta^2}\right), \, \, G_5=-\frac{\mu-4Q_T}{\sqrt{2} H}\cdot \epsilon X^{-1/2}.\end{equation}
Substituting potentials $G_i$ into terms $\mathcal{L}_i$ (\ref{lagr2}), we obtain the equalities:
$$\mathcal{L}_2=G_2=-H^2\left(\frac{3-\beta}{2A}+\mu+(\beta-1)4Q_T\right),$$
\begin{equation}\mathcal{L}_3 =(-1/A+8Q_T)\frac{(3-\beta)^2 H^2}{2\beta}, \, \, \mathcal{L}_5=-8H^2(\mu-4Q_T).\end{equation}
For the estimate we will use dimensionless ratios:
$$\left|\frac{\mathcal{L}_2}{\mathcal{L}_4}\right|=\frac{2Q_T}{3\mu}\left|\frac{3-\beta}{8AQ_T}+\beta+\frac{\mu}{4Q_T}-1\right|,$$
$$\left|\frac{\mathcal{L}_3}{\mathcal{L}_4}\right|=\left|1-(8AQ_T)^{-1}\right|\cdot\frac{(3-\beta)^2 }{|\beta|}\cdot\frac{2Q_T}{3\mu},$$
\begin{equation}\left|\frac{\mathcal{L}_5}{\mathcal{L}_4}\right|=\frac{4}{3}\left|1-\frac{4Q_T}{\mu}\right|.\end{equation}
where we took into account (\ref{qwe}). According to (\ref{3odm}), it takes place
\begin{equation}\label{3odm23} 8AQ_T=c^2_T+c^2_S\cdot\frac{r}{16}.\end{equation}
Next we will consider two cases.

In case $c^2_S\cdot\frac{r}{16}\gg 1-c^2_T>0$, we get the approximation (\ref{doli}), (\ref{b2jw}) for  $\frac{Q_T}{\mu}$ and $\beta$.
Therefore, the relative contributions are of the order
\begin{equation}\label{cont1}\left|\frac{\mathcal{L}_{2,3}}{\mathcal{L}_4}\right|\approx a^{(2,3)}_0+a^{(2,3)}_1(1-c^2_T),\,\, \left|\frac{\mathcal{L}_5}{\mathcal{L}_4}\right|\approx a_2(1-c^2_T),\end{equation}
where $a_i=a_i(r, c_S^2)$. As we can see, the main contribution to the action density is made by $\mathcal{L}_2$, $\mathcal{L}_3$, $\mathcal{L}_4$.

In case $c^2_S\cdot\frac{r}{16}=k^2(1-c_T^2)\ll 1$ ($k\neq 1$), $|1-k^2|(1-c_T^2)\ll r$, we get the approximation (\ref{doem}), (\ref{doem1}) for  $\frac{Q_T}{\mu}$ and $\beta$. Therefore, the relative contributions are of the order
\begin{equation}\label{cont2}\left|\frac{\mathcal{L}_2}{\mathcal{L}_4}\right|=\mathcal{O}(1),\,\, \left|\frac{\mathcal{L}_3}{\mathcal{L}_4}\right|=\mathcal{O}(r),\,\,
\left|\frac{\mathcal{L}_5}{\mathcal{L}_4}\right|=\mathcal{O}((1-c^2_T)^2).\end{equation}
 As we can see, the main contribution to the action density is made by $\mathcal{L}_2$, $\mathcal{L}_3$, $\mathcal{L}_4$.

According to observations, the speed $c_T\approx 1$ is close to unity. The estimate $c_T\approx 1$ gives a factor in $\mathcal{L}_5=-8H^2(\mu-4Q_T) \sim\mu H^2 (1-c^2_T)^\alpha$ ($\alpha=1,\, 2$) and this means that the non-minimal coupling $G_5$ gives the smallest contribution (see (\ref{cont1}), (\ref{cont2})). It is clear that with the exact equality  $c_T=1$  the non-minimal coupling $G_5$ disappears. From (\ref{csT}) and (\ref{w1w4}) the converse statement follows: $G_{5X}=0 \Rightarrow c_T=1$. The other Lagrange functions $\mathcal{L}_2$, $\mathcal{L}_3$, $\mathcal{L}_4$ are non-zero at  $c_T=1$. Thus, the non-minimal coupling $G_5$ ensures that the model matches the observed data on the sound speed  $c_T$.

\section{Conclusion}

We presented the designer  method for flat Friedman–Robertson Walker space-times within the framework of Horndeski's scalar-tensor theory with potentials:
\begin{equation}G_4=\mu/2, \,\,G_2(X,\phi),\,\, G_3(X,\phi),\,\, G_5(X)\neq 0.\end{equation}
 The key ansatz of the method is the assumption of constant values of the perturbation characteristics (the speeds $c_T$, $c_S$, the tensor-to-scalar ratio $r$ and other), which allows us to exclude at the initial stage ghost and Laplace instabilities on all time scales. The end result of applying this method is subclasses of Horndeski's theory that allow cosmological solutions without physical instabilities.As an example, we found subclasses of Horndeski's theory for exponential and power-law inflation models of the Universe without  ghost and Laplace instabilities.

These subclasses have a collective form
$$G_2=S(\phi)\cdot X+A\cdot X^n-V(\phi),$$
\begin{equation}V =\text{const}\cdot \phi^q,\,\,\text{or} \,\, V =\text{const}\cdot e^{-\nu\cdot\phi},\end{equation}
\begin{equation}G_3=B\cdot X^l+G(\phi),\,\,\text{or} \,\,  G_3=C\cdot\ln X+D\cdot\phi,\end{equation}
\begin{equation}G_5=E\cdot X^m,\end{equation}
where $S(\phi)$  is the arbitrary function; $A$, $B$, $C$, $D$, $E$ are constant that are expressed through the perturbation parameters and $\mu$.
 The resulting subclasses contain different field theories. The potential $V(\phi)$ can describe a massive scalar field ($V(\phi)\sim \phi^2$), and can have an exponential form ($V(\phi)\sim exp(-\nu\cdot\phi)$). The function $G_2(X,\phi)$ can correspond to the conventional scalar field ($G_2(X,\phi)=X-V(\phi)$), or $G_2(X,\phi)\sim X^{1/2}$, which corresponds to Cuscuton scenarios \cite{Afshord}. The form $G_3\sim \ln X$ is associated with the existence of the black hole solution with a scalar hairy \cite{Qiong,Tatt}. The logarithmic form $G_3$ was also studied in works \cite{Deffayet,Pujolas}.

 The proposed method contains redundancy of ansatzes, which leads to an equation connecting constant parameters $c_S$, $c_T$, $r$ and others. This means that there are many parameter values for which the condition for avoidance of ghost and Laplacian instabilities  will contradict this equation. Are there parameter values for which there are no contradictions? Such values exist.
For example, observational data $1-3\cdot 10^{-15}<c_T<1$, $0<r<0.1$ do not contradict the equation. Also, the degree $n$ in the scale factor $a(t)\sim t^n$ does not contradict the accelerated expansion of the Universe ($n>1$). Using the de Sitter world as an example, we estimated the contributions of $\mathcal{L}_i$ ($i=2,3,5$) relative to the Einstein-Hilbert term
$\mathcal{L}_4=\mu R/2$ in the action density (\ref{lagr1}). The estimate $c_T\approx 1$ leads to small contribution of non-minimal coupling $G_5$ relative to the Einstein-Hilbert term in the action density. The remaining Horndeski potentials give a contribution comparable to the Einstein-Hilbert term. The non-minimal coupling $G_5$ ensures that the model matches the observed data on the sound speed  $c_T$.

\begin{acknowledgements}
This work was funded by a grant from the Academy of Sciences of the Republic of Tatarstan provided to higher education institutions, scientific and other organizations to support human resource development plans in terms of encouraging their research and academic staff to defend doctoral dissertations and conduct research activities (Agreement No. 12/2025-PD-KFU dated December 22, 2025).
\end{acknowledgements}


\begin{thebibliography}{99}
\bibitem{Horndeski} G.W. Horndeski: Int. J. Theor. Phys. {\bf 10},  363 (1974)
\bibitem{Kobayashi1} T. Kobayashi, M. Yamaguchi and J. Yokoyama: Prog. Theor. Phys. {\bf 126}, 511 (2011)
\bibitem{Gao} C. Gao, J. Cosmol. Astropart. Phys. {\bf 06} (2010) 023.
\bibitem{Granda} L. N. Granda and W. Cardona, J. Cosmol. Astropart. Phys. {\bf 07} (2010) 021.
\bibitem{Shafi1} Q. Shafi and C. Wetterich, Phys. Lett. B {\bf 152} (1985) 51
\bibitem{Shafi2} Q. Shafi and C. Wetterich, Nucl. Phys. B {\bf 289} (1987) 787
\bibitem{Amendola} L. Amendola, Phys.Lett. B 301, 175 (1993)
\bibitem{Sush2012} S.V. Sushkov, Phys. Rev. {\bf D 85} (2012) 123520
\bibitem{Sush2023}S.V. Sushkov, Phys. Rev. {\bf D 108} (2023) 044028
\bibitem{LIGO} Gravitational waves and gamma-rays from a binary neutron star merger: GW170817 and GRB 170817A / LIGO Scientific Collaboration, Virgo Collaboration, Fermi Gamma-Ray Burst Monitor, INTEGRAL // APJ Lett.-- 2017. -- V.848.-- P.L13
\bibitem{Kase} R. Kase and S. Tsujikawa, Int. J. Mod. Phys. D28 (2019) no.05, 1942005
\bibitem{Benjamin} B. Stolzne et al, Astronomy  Astrophysics 707, A323 (2026) [arXiv:2512.11039]
\bibitem{Gu} Gu, G., Wang, X., Wang, Y. et al.  Nat Astron {\bf 9}, 1879–1889 (2025)
\bibitem{Eric18} E.V. Linder, JCAP 1803, 005 (2018)
\bibitem{Eric20} E.V. Linder,  	JCAP 2010, 042 (2020)

\bibitem{DESI1} DESI Collaboration, B. Abareshi  et al., AJ {\bf 164}, 207 (2022)
\bibitem{Hinshaw} G. Hinshaw et al., Astrophys. J. Suppl. Ser. {\bf 208}, 19 (2013)
\bibitem{PAR} P. A. R. Ade et al., Astron. Astrophys. {\bf 594}, A16 (2016)
\bibitem{DESI2} DESI Collaboration, A. G. Adame et al., arXiv e-prints (2024) arXiv:2404.03002
\bibitem{Ellise} G. F. R. Ellis and M. S. Madsen, Class. Quant. Grav. {\bf 8}, 667-676 (1991)
\bibitem{Chervon1} S. V. Chervon and V. M. Zhuravlev, Russ. Phys. J. {\bf 39}, 776-780 (1996); Izv. Vuz. Fiz. {\bf 39N8}, 83-88 (1996)
\bibitem{Chervon2} S. V. Chervon, V. M. Zhuravlev and V. K. Shchigolev, Phys. Lett. B {\bf 398}, 269-273 (1997)
\bibitem{Starobinskyaadf} A. A. Starobinsky, JETP Lett. 68, 757-763 (1998)
\bibitem{Huterer} D. Huterer and M. S. Turner, Phys. Rev. D 60, 081301 (1999)
\bibitem{Nakamura}  T. Nakamura and T. Chiba, Mon. Not. Roy. Astron. Soc. 306, 696-700 (1999)
\bibitem{Chervon3} Fomin, I.V., Chervon, S.V., Eur. Phys. J. C 78, 918 (2018)
\bibitem{Gao1} Yunhao Gao, Baojiu Li, Jie Wang, Nonlinear reconstruction of general dark energy theories,  arXiv:2507.01442v2
\bibitem{Kennedy} J. Kennedy, L. Lombriser, A. Taylor, Phys. Rev. D 96, 084051 (2017)
\bibitem{Arjona} R. Arjona, W. Cardona and S. Nesseris, Phys. Rev. D 100, 063526 (2019)
\bibitem{Bernardo1}  R. C. Bernardo and J. L. Said: JCAP {\bf 09}, 014 (2021)
\bibitem{Bernardo2}  R. C. Bernardo, D. Grandon, J. L. Said and V. H. Cardenas: Phys. Dark Universe
{\bf 36}, 101017 (2022)

\bibitem{Muharlyamov1}  R. K. Muharlyamov,  T. N. Pankratyeva, Eur. Phys. J. Plus {\bf 136}, 590 (2021)
\bibitem{Muharlyamov2}  R. K. Muharlyamov,  T. N. Pankratyeva, Mod. Phys. Lett. A {\bf 37}, 2250108 (2022)
\bibitem{Muharlyamov4} R. K. Muharlyamov, T. N. Pankratyeva, S. O. A. Bashir, Mod. Phys. Lett. A, Vol. 39, No. 17n18, 2450085 (2024)
\bibitem{Muharlyamov3}  R. K. Muharlyamov,  T. N. Pankratyeva,  Indian J. Phys., {\bf 97}, 2239–2245 (2023)
\bibitem{Muharlyamov5} R. K. Muharlyamov, T. N. Pankratyeva, S. O. A. Bashir, Chinese Physics C, Vol. 48, No. 11 (2024) 115107
\bibitem{DeFelice1} A. De Felice and S. Tsujikawa: JCAP {\bf 1202}, 007 (2012)
\bibitem{SushkovStar}  R. Galeev, R. K. Muharlyamov, A. A. Starobinsky, S. V. Sushkov and M. S. Volkov, Phys. Rev. D, {\bf  103}, 104015 (2021)

\bibitem{Charmousis} C. Charmousis, E.J. Copeland, A. Padilla and P.M. Saffin, Phys. Rev. D, {\bf 85}: 104040 (2012)
\bibitem{Bernardo3}  R.C. Bernardo,   	JCAP 03 (2021) 079
\bibitem{SAppleby} S. Appleby, Eric V. Linder,  	JCAP 1807, 034 (2018)
\bibitem{Bernardo4} R.C. Bernardo,  J. L. Said, M. Caruana, S. Appleby,  	JCAP 10 (2021) 078
\bibitem{Bernardo5} R.C. Bernardo,  J. L. Said, M. Caruana, S. Appleby, 2022 Class. Quantum Grav. {\bf 39} 015013
\bibitem{Aghanim} Planck collaboration, Planck 2018 results. VI. Cosmological parameters, Astron. Astrophys. 641
(2020) A6 [Erratum ibid. 652 (2021) C4] [arXiv:1807.06209]
\bibitem{Guth} A. H. Guth, Phys. Rev. D 23, 347-356 (1981)
\bibitem{Linde} A. D. Linde, Phys. Lett. B 108, 389-393 (1982)
\bibitem{Sato} K. Sato, Phys. Lett. B 91, 66 (1981)
\bibitem{Sato1} K. Sato, Mon. Not. R. Astron. Soc. {\bf 195}, 467 (1981)
\bibitem{Belinsky} V. A. Belinsky, I. M. Khalatnikov, L. P. Grishchuk and Y.B. Zeldovich, Phys. Lett. B, {\bf 155}, 232-236 (1985)
\bibitem{Piran} T. Piran, Tsvi and W. M. Ruth, Phys. Lett. B {\bf 163}, 331-335 (1985)
\bibitem{Piran1} T. Piran, Phys. Lett. B {\bf 181}, 238-243 (1986)
\bibitem{Halliwell} J. J. Halliwell, Phys. Lett. B {\bf 185}, 341 (1987)
\bibitem{Abbott} L. F. Abbott and M. B. Wise, Nucl. Phys. B {\bf 244}, 541 (1984)
\bibitem{Lucchin} F. Lucchin and S. Matarrese, Phys.Rev. D 32, 1316 (1985)
\bibitem{Barrow1} J. D. Barrow, Phys. Lett. B 187, 12-16 (1987)
\bibitem{Muller} V. Muller, H. J. Schmidt and A. A. Starobinsky, Class. Quant. Grav. 7, 1163-1168 (1990)
\bibitem{Afshord} N. Afshordi, D. J. H. Chung and G. Geshnizjani, Phys. Rev. {\bf D 75}, 083513 (2007)
\bibitem{Qiong} Q.Fang, S.Chen and J. Jing, Int. J. Mod. Phys. D, {\bf 28}: 1950112 (2019)
\bibitem{Tatt} O. Tattersal, P. Ferreira, M. Lagos, Phys. Rev. D, {\bf 97}: 084005 (2018)
\bibitem{Deffayet} C. Deffayet, O. Pujolas, I. Sawicki, A. Vikman, J. Cosmol. Astropart. Phys., {\bf 10}: 026 (2010)
\bibitem{Pujolas} O. Pujolas, I. Sawicki, A. Vikman, J. High Energy Phys. {\bf 11}: 156 (2011)







\end{thebibliography}
\end{document}